\documentclass[aps,pra,twocolumn,groupedaddress,showpacs]{revtex4}
\usepackage{amsmath,amssymb}
\usepackage[dvips]{graphicx}

\begin{document}
\title{Rotating Morse wave packet dynamics of diatomic molecule}

\author{Utpal Roy \footnote{e-mail:
utpal.roy@unicam.it, utpalphys@gmail.com}}

\author{Suranjana Ghosh \footnote{e-mail:
suranjana.ghosh@unicam.it}}

\affiliation{Department of Physics, University of Camerino,
I-62032 Camerino, Italy}

\begin{abstract}
We investigate the dynamics of a rotating Morse wave packet,
appropriate for a ro-vibrating diatomic molecule. The coupling
between vibrational and rotational degrees of freedom is
explicated in real position space as well as in phase space Wigner
distribution of a SU(2) coherent state at various dynamically
evolved times. We choose the well studied $I_{2}$ molecule with
the parameter values in good agrement with experiments. A
quantitative measure of the angles of rotations for different
angular momenta is also given.

\end{abstract}

 \maketitle

\section{Introduction}
The Morse potential \cite{morse} is extensively studied in the
literature and has been a subject of considerable attention. It
involves the vibrational degrees of freedom of a diatomic
molecule. On the other hand, rigid rotor model is described to
take into account only the rotational degrees of freedom, where
on-axis rotation is usually neglected and end-over-end rotations
are identical for a homo-nuclear molecule. The rotational and
vibrational energy scales are far apart and it has been a
customary to treat them independently. However, in reality, this
separation is inadequate and the internuclear distance is affected
by the centrifugal force resulting from rotation. Thus, one should
consider both degrees of freedom of a diatomic molecule. A
combined form of the Morse potential and the rotational
centrifugal term is known as the rotating Morse potential. It
leads a more complex energy spectrum reflecting the coupling among
the two intrinsic properties of nuclear molecule. It is known that
the Schr\"{o}dinger equation for this potential can be exactly
solved only when the orbital angular momentum quantum number $j$
is equal to zero. Since, there is no exact analytical solution for
the rotating Morse potential in general, some semiclassical and/or
numerical solutions have been obtained by using various
approximate methods. The original approach developed by Pekeris
\cite{pekeris} is based on the expansion of the centrifugal
barrier in a series of exponentials depending on the internuclear
distance up to second order. A more appropriate expansion has been
proposed by Duff and Rabitz \cite{rabitz} and Elsum and Gordon
\cite{gordon}. These semi-analytical models are based on the idea
of building up Morse potentials with parameters dependent on the
rotational quantum number, which represent adequately the
effective potential curves of the rotating Morse oscillator. There
are various other numerical and proposed methods to study the
rotating Morse potential
\cite{filho,morales,kill,bag,berk,bayrak,qiang,nasser,jose}. These
studies are very useful to obtain more accurate energy eigen
values.

Recent development of ultrashort pulses opened a new research
field which allows to study the wave packet dynamics of diatomic
molecule in real time scales. Pulses of a few tens of femtosecond
allow a coherent broadband excitation with preparation and
detection of rovibrational dynamics. In all previous works, the
wave packet dynamics have been studied in diatomic molecule either
considering the vibration \cite{vetchin,stolow,garraway} and
rotation \cite{seideman,comstock,spanner} separately. There are
few works, which considers the both in two-time scale
configuration \cite{lohm,suranjana,cao}.

In this work, we study the wave packet dynamics, considering the
coupling among the vibrational and rotational degrees of freedom
of nuclear motion. Our analytical approach for the study of the
rotating Morse potential is inspired by Gordon's method
\cite{gordon}, where the effective potential becomes minimum
around a new equilibrium internuclear distance $r_{j}$, which is a
function of angular momentum quantum number $j$. This model, so
far, most appropriately manifests the coupling between two degrees
of freedom, vibration and rotation. To the authors' knowledge,
there is no study about the wave packet dynamics of this
appropriate system. We show that this $1$D model is accurate upto
a reasonably high angular momentum ($j=200$), accessible to
experiments. We study rovibrational interplay through a coherent
state (CS) dynamics, not only in coordinate space, but also in
phase space Wigner distribution. It is shown that rovibrational
interplay is better explicated in phase space rotation of the CS
for different angular momenta. When rovibrational dynamics is seen
in $3$D configuration space, the corresponding phase space is six
dimensional, which is not possible to visualize. Avoiding this
$3$D intricacy, this $1$D rotating Morse potential can well
describe the coupling between the two degrees of freedom, hence
can nicely manifest the rovibrational interplay for this realistic
system in phase space. It also shows the effects of
rotation-vibration in the quantum interference regime. We have
chosen the iodine molecule $(I_{2})$ which is an unique suited
seed molecule for laser induced fluorescence. Additionally, we
give a numerical method, which can estimate the angles of
rotation, are in good agreement with the Wigner plots.

\begin{figure*}[htpb]\centering
\includegraphics[width=4.8in]{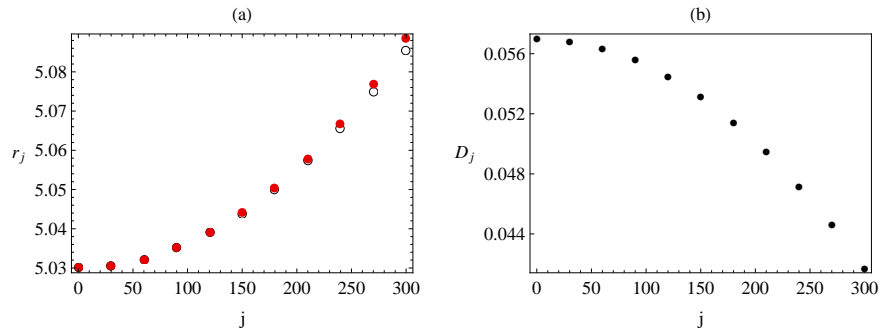}
\caption{(a) The variation of $r_{j}$ with the rotational quantum
number $j$. It implies that the numerical values, obtained by
solving the transcendental Eq.~(\ref{transcendental}) [red filled
circles in (a)] matches nicely with the approximate values from
Eq.~(\ref{rjdj}) [black circles in (a)]. It starts to differ for
higher values of $j$ ($>200$). The change of the dissociation
energy of the effective potential with $j$ is depicted in (b).}
\label{rjdj2D}
\end{figure*}

The paper is organized as follows. In the following section, we
present a brief overview and justification of the $1$D rotating
Morse system. We also evaluate the corresponding SU(2) CS wave
packet. In Sec. III, we analyze the dynamics of the CS of a
diatomic molecule and explain the effect of rovibrational coupling
in this. After showing the effect in the configuration space, we
move to the phase space picture for a better representation, which
clearly displays the rotational effect. A quantitative analysis of
the angle of rotation is also given. This numerical analysis
confirms the values of rotation angle, corresponding to various
angular momentum quantum number. Finally, we end with some
conclusions in Sec. IV.

\section{Rotating Morse wave packet}

The Schr\"{o}dinger equation for the rotating Morse oscillator
reads as follows,
\begin{equation}\label{rotmorse}
\left[-\frac{\hbar^2}{2\mu}\frac{d^2}{d
r^2}+V_{e\!f\!f}(r)\right]\psi_{n,j}(r)=E_{n,j}(r),
\end{equation}
where $\mu$ is the reduced mass of the diatomic molecule and $n$,
$j$ are the vibrational and rotational quantum numbers,
respectively. The effective potential $V_{e\!f\!f}(r)$ has the
form
\begin{equation}\label{effpot}
V_{e\!f\!f}(r)=D[e^{-2\beta(r-r_{0})}-2e^{-\beta(r-r_{0})}]+\frac{j(j+1)\hbar^2}{2\mu
r^2}.
\end{equation}
The first part describes the well known Morse potential, where $D$
is the dissociation energy of the Morse potential, $r_{0}$ is the
equilibrium internuclear separation and $\beta$ is the range
parameter. The second part of the above equation stands for the
centrifugal contribution of rotation. More accurate description of
the system can be obtained with a modified equilibrium
internuclear distance $r_{j}$ and a dissociation energy $D_{j}$
\cite{gordon}. Using a semi-analytical method \cite{burkhardt},
one can find
\begin{equation}\label{rjdj}
r_{j}=r_{0}\left[1+\frac{A}{\beta^2r_{0}^2D}\right];\;\;\;D_{j}=D\!-\!A\left(1-\frac{A}{\beta^{2}r_{0}^{2}D}\right),
\end{equation}
where $A\!=\!\frac{j(j+1)\hbar^2}{2\mu r_{0}^2}$. Alternatively,
$r_{j}$ can also be computed numerically by solving the
transcendental equation
\begin{equation}\label{transcendental}
\frac{dV_{e\!f\!f}(r)}{dr}\mid_{r=r_j}=0.
\end{equation}
These two sets of $r_{j}$'s are plotted in Fig.~\ref{rjdj2D}(a),
which shows very good agrement for $j<200$. It shows that the
rotational potency results into the increasing value of the
equilibrium distance [Fig.~\ref{rjdj2D}(a)] and decreasing nature
of dissociation energy [Fig.~\ref{rjdj2D}(b)]. Physically, in
presence of the rotational centrifugal force, the two constituent
atoms of a diatomic molecule tend to settle in a larger distance
and they are more prone to dissociate, reducing the amount of
energy required to make them independent. We define
$A_{j}\!=\!\frac{j(j+1)\hbar^2}{2\mu r_{j}^2}$ and expand the
centrifugal term of Eq.(\ref{effpot}) around $r=r_{j}$. Keeping
terms upto second order, the effective potential becomes
\begin{equation}
V_{e\!f\!f}(y)=\frac{c_{2}}{4\lambda_{j}^2}y^2-\frac{c_{1}}{\lambda_{j}}y+c_{0},
\end{equation}
where $y\!\!=\!\!2\lambda_{j} e^{-\beta
(r-r_{0})}\!\!=\!\!2\lambda_{j} u e^{-\beta (r-r_{j})}$ and
$u\!\!=\!\!e^{-\beta (r_{j}-r_{0})}$. The parameter,
$\lambda_{j}=\sqrt{\frac{2\mu c_{2}}{\beta^2\hbar^2}}$, where the
constants, $c_{0}$, $c_{1}$, and $c_{2}$ are dependent on quantum
number $j$:
\begin{eqnarray}
c_{0}=3A_{j}b^{2}_{j}-3A_{j}b_{j}+A_{j},\nonumber\\
c_{1}=(3A_{j}b^{2}_{j}-2A_{j}b_{j}+A_{j}+u D)/u,\nonumber\\
and\;\;  c_{2}=(3A_{j}b^{2}_{j}-A_{j}b_{j}+A_{j}+u^2D)/u^2,
\end{eqnarray}
with $b_{j}\!=\!(\beta r_{j})^{-1}$. Under this approximation, the
Schr\"{o}dinger equation (\ref{rotmorse}) is solved to yield the
eigen functions of this rotating Morse system:
\begin{equation}\label{eigenstate}
\psi_{n,j}(y)= N_{n,j} e^{-y/2} y^{s} L_{n}^{2s} (y),
\end{equation}
where the variable $y$: $0<y<\infty$, $n$ is the vibrational
quantum number, $L_{n}^{2s} (y)$ stands for associate Laguerre
polynomial and
$s=\sqrt{-\frac{(E_{v,j}-c_{0})\lambda^2_{j}}{c_{2}}}$. The number
of bound states in the system is $[\bar{\lambda_{j}}-1/2]$, where
$[\rho]$ denotes the largest integer smaller than $\rho$. The
parameter $\bar{\lambda}_{j}$ is related as
$\bar{\lambda}_{j}=\frac{c_{1}}{c_{2}}\lambda_{j}$.

They satisfy the constraint condition $s+2n=2\bar{\lambda}_{j}-1$.
$N_{n,j}$ is the normalization constant:
\begin{equation}
N_{n,j}\!=\!\left[\frac{\beta(2\bar{\lambda}_{j}-2n-1)\Gamma{(n+1)}}{\Gamma{(2\bar{\lambda}_{j}-n)}}\right]^{1/2}.
\end{equation}
The rovibrational energy eigen values $E_{n,j}$ turns out as
\begin{equation}
E_{n,j}=2\frac{c_{1}}{\lambda_{j}}(n+1/2)-\frac{c_{2}}{\lambda_{j}}(n+1/2)^2+c_{0}-\frac{c_{1}^2}{c_{2}}.
\end{equation}
It is worth pointing out that in absence of rotation
$c_{0}\!=\!0$, $c_{1}\!=\!c_{2}\!=\!D$, the system only describes
a vibrating diatomic molecule.

\section{Dynamics of a rotating Morse wave packet of a diatomic molecule}
Ultrashort laser pulses can probe the molecular dynamics in real
time scale. Time-resolved measurements on the femtosecond scale
have made it possible to know the evolution of a coherently
prepared superposition of ro-vibrational states. Here, we consider
a rovibrational wave packet of $I_{2}$ molecule, which is a
coherent state, dependent on particular rotational quantum number.
Many theoretical and experimental investigations have been carried
out on this molecule \cite{zewail1,lohm,fischer} and has been a
subject of many femtosecond studies, particularly in wave packet
interferometer \cite{ohmri,ohmri09}. The corresponding parameter
values are $\beta=0.9849\;a.u.$, reduced mass
$\mu=11.56\times10^4\;a.u.$, $r_{0}=5.03\;a.u.$, and
$D=0.057\;a.u.$. In particular, Zewail and co-workers investigated
rovibrational wave packet dynamics in the well-characterized
electronic $B0^{+}_{u}$ state \cite{zewail1}. Lohm\"{u}ller
\emph{et al.} \cite{lohm} discussed about the pump-probe
experiment of $I_{2}$ in a room temperature and the detection of
fractional revivals using full-dimensional quantum wave packet.
Following Lohm\"{u}ller \emph{et al.}, we consider an initial
rovibrational wave packet centered around $10$th vibrational
energy level and for fixed $j=60$. This can be excited by using a
$620$ nm pump pulse, which is in good agreement with the
experiments \cite{zewail1,zewail2} and for detection purpose,
probe pulses of $310$ nm and $330$ nm are used. These parameters
allow the probe excitation mainly between the $B0^{+}_{u}$ and
$f0^{+}_{g}$ states \cite{wheeler}. Under the laser polarizations
magic angle conditions, it takes into account the vibrational as
well as the rotational motions.

\begin{figure}[htpb]
\begin{center}
\includegraphics[width=3.in]{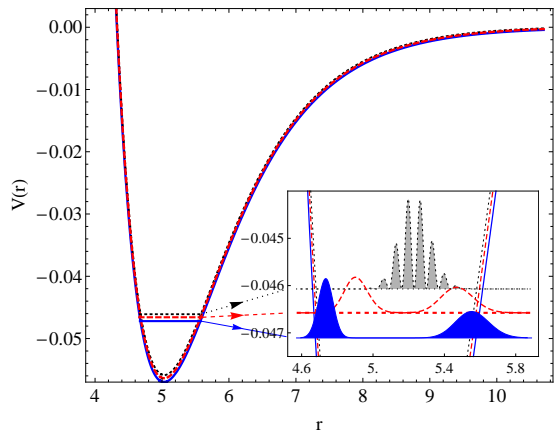}
\caption{Effective potentials for $j=0$ (solid line), $j=60$
(dashed line) and $j=81$ (dotted line) are depicted. The
corresponding $10$th energy levels are zoomed in the inset plot.
The wave packets at $t=0.25\;T_{rev}$ are also shown in the inset
as dark filled ($j=0$), dashed ($j=60$) and light filled ($j=81$)
plots, respectively. Changes of $j$ value cause slight changes in
the effective potential, which result in rotational effect on the
wave packet. The two splitted wave packets in co-ordinate space
come closer with increasing $j$ and superpose each other for
$j=81$, producing interference ripples. The potential and the
internuclear distance are in atomic units.} \label{2D}
\end{center}
\end{figure}

The rotational effect on the effective potential of $I_2$ molecule
is depicted in Fig.~\ref{2D}. Without rotation \emph{i.e.}, $j=0$
(solid line), it is nothing but the Morse potential. Once rotation
is considered, e.g., $j=60$ (dashed line) or $j=81$ (dotted line),
the potential changes, i.e., the potential-minima slowly shifts
towards right (from $r_e = 5.03$ a.u. to $r_j =5.0323$ a.u. and
$r_j = 5.034$ a.u. respectively) and dissociation energy
diminishes. The rovibratational coupling changes the effective
potential and hence energy levels are shifted upwards. The $10$th
energy level, where the coherent state is excited, is shifted from
$E_{10}=-0.04719$ a.u. to $E_{10}=-0.046595$ a.u. and
$E_{10}=-0.046108$ a.u respectively. In our further study, in
particular, we consider these effective potentials and the initial
wave packet as a coherent state involving few lower levels of
$I_{2}$ molecule centered around $n=10$. If we take only bound
states of the potential, the dynamical symmetry group is $SU(2)$.
In this case, the displacement operator coherent state can be
constructed and written as
\begin{equation}
\Phi(y,t)=\sum_{n=0}^{n'}d^{j}_{n} \;\psi_{n,j}(y) e^{-iE_{n,j}t}.
\end{equation}

\begin{figure*}
\centering
\includegraphics[width=5.7in]{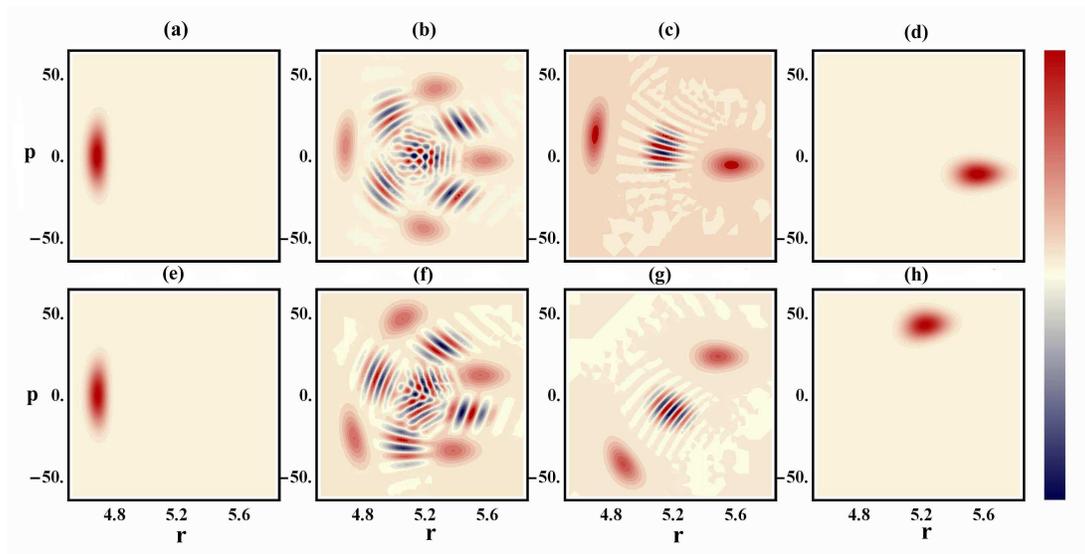}
\caption{(Color online) Time evolution of Wigner function of the
CS for $j=0$ at times (a) $t=0$, (b) $t=T_{\mathrm{rev}}/8$, (c)
$t=T_{\mathrm{rev}}/4$, and (d) $t=T_{\mathrm{rev}}/2$. The second
row shows the corresponding plots for $j=60$ [see (e)-(h)]. These
structures $(j=60)$ reveal the different rovibrational coupling at
different evolved times. Position and momentum variables, $r$ and
$p$, are in atomic units.}\label{wig1}
\end{figure*}

\begin{figure}
%\centering
\includegraphics[width=3.5in]{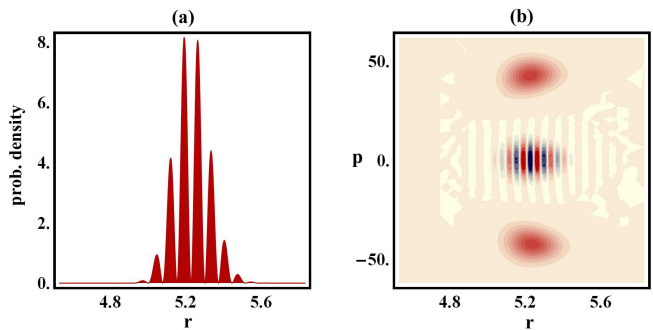}
\caption{(Color online) (a) Probability density of rotating Morse
wave packet for $j=81$ at $t=T_{\mathrm{rev}}/4$. (b) The
corresponding phase space Wigner distribution at the same time.
Position and momentum variables, $r$ and $p$, are in atomic
units.}\label{wig2}
\end{figure}

Here, the group generators are $j$ dependent and the weighting
coefficient becomes
\begin{equation}
d^{j}_{n}=\frac{(-\alpha)^{n'-n}}{(n'-n)!}\left[\frac{n'!
\Gamma(2\bar{\lambda}_{j}-n)}
{n!\Gamma(2\bar{\lambda}_{j}-n')}\right]^\frac{1}{2}.
\end{equation}
It can be easily verified that for $j=0$, the above coherent state
(CS) is the same as reference \cite{ghosh1}. The $SU(2)$
generators satisfy the commutator relations,
$[\hat{J_{+}},\hat{J_{-}}]=2\hat{J_{0}}$,\;$[\hat{J_{0}},\hat{J_{\pm}}]=\pm\hat{J_{\pm}}$
and also satisfy
\begin{equation}
[\hat{J_{+}},\hat{J_{-}}]\psi_{n,j}(y)=2j_{0}\psi_{n,j}(y),
\end{equation}
at the level of wave function, where
$j_{0}=n-\bar{\lambda_{j}}+1/2$. $\hat{J_{0}}$ is the projection
operator of the angular momentum $m$: $m=n-\bar{\lambda_{j}}+1/2$.
Although, the CS of a physical system is never ideal, like that of
harmonic oscillator, we find that our chosen CS is very well
localized and remains coherent for a reasonable time duration.
Presence of quadratic term in energy expression, leads some
interesting phenomena, called fractional revivals, which occur at
some specific instances between two full revivals ($T_{rev}$)
\cite{averbukh,robinett}. The short time evolution displays a
classical periodicity ($T_{cl}$). The classical and revival time
periods are respectively given by
\begin{equation}
T_{cl}=\frac{2\pi\lambda_j}{2c_{1}\!-\!c_{2}\lambda_j}\;\;\; and
\;\;\; T_{r\!ev}=2\pi\lambda_j^{2}/c_{2}.
\end{equation}
For $I_2$ molecule, these time scales are $T_{cl}=0.156$
picosecond and $T_{rev}=36.2$ picosecond. At fractional revival
times (rational fraction ($r/s$) of the revival time $T_{r\!ev}$),
the wave packet breaks into several number of subsidiary wave
packets, where $r$ and $s$ are mutually prime integers. For even
values of $s$ the wave packet breaks into $s/2$ in number,
otherwise breaks into $s$ parts. In the course of time evolution,
one obtains the Schr\"{o}dinger cat like state at one fourth of
the revival time. Four-way break up or the compass-like state
emerges at one-eight of the revival time. Latter produces
sub-Planck scale structures in phase space Wigner distribution. In
the inset of Fig.~\ref{2D}, the $10$th energy levels for different
angular momenta are zoomed and the rotational coupling effect is
shown at $t=0.25\;T_{rev}$, when CS is splitted into two parts.
Initially ($j=0$), the two parts are situated at $r=4.7$ a.u. and
$5.58$ a.u. (dark filled plot). After the introduction of
rotational coupling ($j=60$), they come close to each other,
situated at $r=4.87$  a.u. and $5.48$  a.u. respectively (dashed
line). For a greater angular momentum ($j=81$), the position space
probability structure looks completely different and shows
oscillatory structure (light filled plot). The interpretation lies
in the fact that both the splitted CSs oscillate inside the
potential well in back and forth motion. In the first quarter of
the oscillation, they approach each other, while in the next
quarter, they recede. At halfway of the oscillation, they reflect
from the potential well with a phase change $\pi$ and become again
counter-propagating. For the angular momentum $j=81$, they overlap
each other in the course of their oscillation and produce the
oscillatory ripples, clearly visible in the inset of Fig.~\ref{2D}
and in Fig.~\ref{wig2}(a).

Until now we have confined our study in position space
explanation. This explanation does not reveal the full description
of rotational coupling of the system. Hence, we enlarge our view
towards the phase space description. Wigner function
\cite{schleich} is very useful tool to visualize the quantum
dynamics in phase space and is defined as
\begin{eqnarray}\label{wigner}
W(r,p,t)&=&\frac{1}{\pi\hbar}\int_{-\infty}^{+\infty}
\Phi^{*}(r-r',t)\nonumber\\&& \times\Phi(r+r',t)
e^{-2ipr'/\hbar}dr'\;.
\end{eqnarray}
\begin{figure}
\includegraphics[width=2.8in]{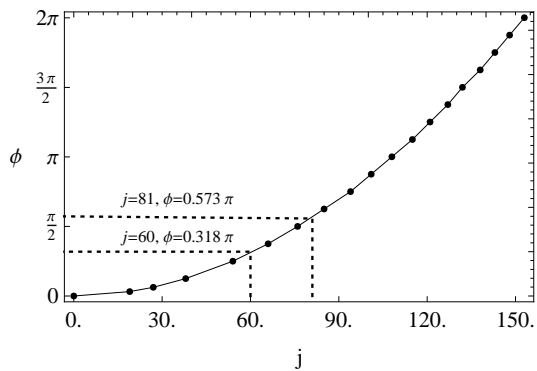}
\caption{Variation of angle of rotation $\phi$ of the wave packet
with the angular momentum quantum number $j$. The dotted lines
specify two particular rotations corresponding to two angular
momenta. $\phi$ is in the unit of radians.}\label{rotation}
\end{figure}
This representation can reveal interesting structures of the CS at
different evolved times. As $I_{2}$ molecules has large number of
bound states $(117)$, we choose the coherent state parameter
$\alpha=2.15$, which involves few lower levels of the potential.
In Fig.~\ref{wig1}, first row shows the time evolution of the CS
with $j=0$. It represents the Morse vibrational dynamics at
different fractional revival times. The initial CS is situated
near the left potential boundary and is well localized
[Fig.~\ref{wig1} (a)]. Owing to the anharmonicity of the
potential, the probability densities near the left boundary are
stiffer compare to those near right boundary. From the previous
explanation, it is evident that the compass-like state appears at
$t=T_{\mathrm{rev}}/8$ [Fig.~\ref{wig1} (b)], which shows the
smallest interference structures \emph{i.e.}, the well known
sub-Planck scale structures \cite{zurek,ghosh1,ghosh2,ghosh3}.
Origin of these smallest structures is the superposition of the
two diagonal interference ripples. The Schr\"{o}dinger cat-state
\cite{foldi} appears at $t=T_{\mathrm{rev}}/4$
[Fig.~\ref{wig1}(c)], where the interference ripples appear in the
middle as alternate positive and negative values, along the line
joining the two component CSs. All these negative phase space
interference structures are the strong signature of quantum
behavior. At half revival time the CS revives near the right
boundary of the potential well [Fig.~\ref{wig1} (d)]. Upon
introducing the angular momentum contribution, we show the phase
space structures at the same times as above in
Fig.~\ref{wig1}(e)-(h). The initial CS is similar because it does
not depend on rotation. One can observe the rotational effect on
the other structures. For heavy molecules, the rotational effect
is less for smaller values of $j$. We choose $j=60$ as it is
reasonable and also appropriate for experiment. Interestingly, the
revival time $T_{rev}$, does not change with $j$ values, because
the ratio $\lambda_j^{2}/c_{2}$ remains same. Whereas, the
classical period $T_{cl}$ alters due to the rovibrational
coupling. The rotation is anti-clockwise, thereby making the angle
of rotation positive in our convention. Fig.~\ref{wig1}(f)
describes the rotational effect at the sub-Planck interference
structures. It is established that these structures are very
sensitive to external perturbation or rotation, which can be
utilized for quantum parameter estimations \cite{toscano}. These
sub-Planck structures are also found most sensitive against
decoherence in diatomic molecular system of hydrogen iodide
molecule \cite{ghosh2}. In our model, it is worth to study the
rotational sensitivity due to a infinitesimal rovibrational
coupling in this physical system. As this is not the present goal
of this work, we are studying it in another context \cite{prep}.
Fig.~\ref{wig1}(g) shows the rotational effect in the cat state.
The CS is revived at $t=0.5\;T_{rev}$, but shifted in its orbit
due to rotation.

We would like to explain another specific case of the phase space
structures to reveal the intricacies in a better way.
Fig.~\ref{wig2} shows the cat state for higher values of $j=81$.
This cat state is separated in momentum space, but exactly
superposed in real position space. A single interference ripple in
Fig.~\ref{wig2}(a) has dimension $~0.07$ a.u. or $3.7$ picometer.
Although the experimental observation of small quantum
interference structures is very challenging, similar interference
ripples in picometer scale is recently visualized experimentally
for $I_2$ molecule \cite{ohmri}. They observed it in the classical
time evolution of a vibrational wave packet at time
$t=0.25\;T_{rev}+0.25\;T_{cl}$ or $t=0.25\;T_{rev}+0.75\;T_{cl}$.
The same picometer structures can be observed for a rovibrating CS
of $I_2$ molecule at a fixed time ($t=0.25\;T_{rev}$) with angular
momentum around $j=81$. This structure remains stationary in their
full crossing pathway in the range $74<j<88$, which keeps the
flexibility of experiments.

Now a natural question can be asked whether there is a way to know
the orientation $\phi$ corresponding to a $j$ value. Here, we
provide a numerical estimation of this rotation angle. We operate
an operator $U=e^{i\hat{J}_{0}\phi}$, where $\hat{J}_{0}$ is the
angular momentum operator, on the initial wave packet and obtain
\begin{eqnarray}\label{angle}
U \Phi(y,t)_{j=0}&=& \sum_{n=0}^{n'}d_{n}^{0}e^{i
(n-\bar{\lambda_{j}}+1/2)\phi}\psi_{n,0}e^{-iE_{n,0}t}\nonumber\\&=&\chi(y,t).
\end{eqnarray}
Then we compute the overlap $|\!<\chi(y,t)|\Phi(y,t)>\!|^2$,
numerically and take the value of $\phi$ for the largest overlap
$\sim 1$. We repeat it for various angular momenta and find the
corresponding angles of rotation. In particular, we have chosen
the rotation of cat state, \emph{i.e.}, at time $0.25\;T_{rev}$.
The result is depicted in Fig.~\ref{rotation} upto a full period
of rotation ($2\pi$ radians). Two particular rotations ($j=60$ and
$j=81$) are indicated in the figure by dotted lines. From the
nature of the plot, one can infer that the wave packet is more
phase sensitive at larger $j$ values. We have checked that these
rotation angles are in good agreement with the phase space
pictures. To be more precise, let us take a particular case for
$j=81$. We obtained an angle of rotation $\phi=0.573 \pi$ radians
for $j=81$, but the cat state is exactly vertically situated
($0.5\pi$ radians) with respect to the $p=0$ line
(Fig.~\ref{wig2}(b)). This is because, the initial cat state (j=0,
Fig.~\ref{wig1}(c)) is not exactly horizontal due to the
$T_{rev}/T_{cl}$ ratio, which is not exactly an integer and
imprinting a small phase or rotation equivalent to
$\phi=-0.073\pi$ radians. Thus, for a perfect vertical cat state,
one needs the rotation of $0.573\pi$ radians, which is exactly the
obtained value in our analysis.

\section{Conclusion}
We have showed the time evolution of a CS wave packet of rotating
Morse potential. We are able to manifest the rovibrational
coupling of a diatomic molecule through this CS dynamics in the
experimentally allowed parameter regime. We considered a well
behaved and appropriate SU(2) coherent state, including few bound
states of $I_2$ molecule. This can be generated with an available
pump beam excited at the $10$th energy level and for angular
momenta $j=60$ or $81$. The rovibrational coupling is explored in
position space, where the physical rotation of the CS is not so
transparent. Phase space structure clearly showed the rotational
influences on the cat state as well as the compass-like state. Our
study deprecates the complication of six-dimensional physics in
phase space for ro-vibrational dynamics. The $1$-D rotating Morse
potential can well capture the rotational effect throughout the
time evolution in phase space. Moreover, we provide a way to
identify the amount of rotation of the rovibrating CS at arbitrary
time and angular momentum, which can be very useful to know the
behavior of the system at certain extend, without going into the
phase space description.

\end{document}